\documentclass[twocolumn]{aa}

\usepackage{graphicx}
\usepackage{amsmath,amsfonts,amssymb}
\usepackage{txfonts}
\usepackage{color}
\usepackage{natbib}
\usepackage{float}
\usepackage{dblfloatfix}
\usepackage{afterpage}
\usepackage{ifthen}
\usepackage[morefloats=12]{morefloats}
\usepackage{placeins}
\usepackage{multicol}
\bibpunct{(}{)}{;}{a}{}{,}
\usepackage[switch]{lineno}
\definecolor{linkcolor}{rgb}{0.6,0,0}
\definecolor{citecolor}{rgb}{0,0,0.75}
\definecolor{urlcolor}{rgb}{0.12,0.46,0.7}
\usepackage[breaklinks, colorlinks, urlcolor=urlcolor,
    linkcolor=linkcolor,citecolor=citecolor,pdfencoding=auto]{hyperref}
\hypersetup{linktocpage}
\usepackage{bold-extra}
\usepackage{xcolor}
\usepackage{algpseudocode}
\usepackage{comment}

\usepackage{lipsum}

\def\Planck{\emph{Planck}}

\providecommand{\sorthelp}[1]{}

\begin{document}

\title{Optimal bolometer transfer function deconvolution for CMB experiments through maximum likelihood mapmaking}
\newcommand{\oslo}[0]{1}

\author{\small
A.~Basyrov\inst{\oslo}\thanks{\url{artem.basyrov@astro.uio.no}}
\and
N.~O.~Stutzer\inst{\oslo}
\and
J.~G.~S.~Lunde\inst{\oslo}
\and
H.~K.~Eriksen\inst{\oslo}
\and
E.~Gjerløw\inst{\oslo}
\and
D.~J.~Watts\inst{\oslo}
\and
I.~K.~Wehus\inst{\oslo}
}
\institute{\small
Institute of Theoretical Astrophysics, University of Oslo, Blindern, Oslo, Norway\goodbreak
}

\authorrunning{Basyrov et al.}
\titlerunning{Optimal transfer function deconvolution}

\abstract{
  We revisit the impact of finite time responses of bolometric detectors used for deep observations of the cosmic microwave background (CMB). Until now, bolometer transfer functions have been accounted for through a two-step procedure by first deconvolving an estimate of their Fourier-space representation from the raw time-ordered data (TOD), and then averaging the deconvolved TOD into pixelized maps. However, for many experiments, including the \Planck\ High Frequency Instrument (HFI), it is necessary to apply an additional low-pass filter to avoid an excessive noise boost, which leads to an asymmetric effective beam. In this paper we demonstrate that this effect can be avoided if the transfer function deconvolution and pixelization operations are performed simultaneously through integrated maximum likelihood mapmaking. The resulting algorithm is structurally identical to the \texttt{artDeco} algorithm introduced by \cite{ArtDeco} for beam deconvolution. We illustrate the relevance of this method with simulated \Planck\ HFI 143\,GHz data, and find that the resulting effective beam is both more symmetric than with the two-step procedure, resulting in a sky-averaged ellipticity that is 64\,\% lower, and an effective beam full-width-at-half-maximum (FWHM) that is 2.3\,\% smaller. Similar improvements are expected for any other bolometer-based CMB experiments with long time constants.
}


\maketitle


\section{Introduction}

During the last three decades our understanding of the cosmic microwave background (CMB) has been revolutionized by a series of increasingly sensitive instruments (e.g., \citealt{COBE_4years, WMAP_9years_cosm, planck2013-p01, planck2014-a01, planck2016-l01}). These advances have been made possible by increased sensitivity, driven by improvements in detector technology both for coherent radiometer and incoherent bolometer detectors.

Bolometric detectors in particular have gone a long way in improving both sensitivity and the frequency range they are able to operate in \citep[e.g.,][]{Zhao_2008, bersanelli2010, lamarre2010, Stevens_2020}. One of the main characteristics of a bolometer is a finite time constant that describes its temporal response to a signal change \citep[e.g.,][]{planck2013-p03c}. The main observational signature of a finite bolometer transfer function is an apparent smoothing of the true underlying signal along the scanning path of the instrument. Fortunately, the magnitude of this effect has diminished over time, as the bolometer detector technology has improved and the response rates have become faster. Still, this effect has to be accounted for during mapmaking in order to establish an accurate estimate of the true sky signal. 

The traditional approach to account for this effect is simply to deconvolve an estimate of the bolometer transfer function from the time-ordered data (TOD), which results in an unbiased signal. A significant drawback of this method, however, is that it not only affects the sky signal, but also the noise measured by the detector. The noise is therefore effectively amplified on short time-scales. In the original \Planck\ analysis, this problem was solved by applying an extra low-pass regularization filter function that suppresses high-frequency noise \citep{planck2013-p03c}. While using this filter does solve the noise amplification problem at high temporal frequencies, it also modifies the signal, thereby introducing an extra component to the effective beam. 

In this work, we propose an alternative approach that exploits the same ideas as proposed for beam deconvolution by \cite{ArtDeco}. Specifically, rather than explicitly deconvolving the beam transfer function in a pre-processing step prior to mapmaking, we integrate the deconvolution operator directly into the maximum likelihood mapmaking equation, which then is solved using a conjugate gradient method (e.g., \citealt{CG_pain}). This approach has several advantages. Firstly, it does not require an explicit additional noise regularization kernel, but relies on the scanning strategy itself to regularize the high-frequency noise. This method yields an  unbiased estimate of the true sky signal without modifying the effective beam. Secondly,  it results in significantly weaker noise correlations at high temporal frequencies. The main drawback of the method is a higher computational expense.

The rest of the paper is organized as follows. We first describe the new method in Sect.~\ref{sec:method}. We then illustrate the main points with a simple one-dimensional case in Sect.~\ref{sec:1D_test} in which all calculations can be performed by using dense linear algebra. Next, we consider the two-dimensional case in Sect.~\ref{sec:point_source}, and start by characterizing its performance for a grid of point sources. Finally, we apply the method on the simulated CMB map in Sect.~\ref{sec:CMB_case} with properties similar to the \Planck\ 143\,GHz channel. We  discuss the computational cost of the proposed method in Sect.~\ref{sec:cost}, before concluding in Sect.~\ref{sec:conclusions}.

\section{Method}
\label{sec:method}

We start by assuming that the data $\vec{d}$ recorded by a bolometer may be modelled as 
\begin{equation}
    \label{eq:TOD}
    \vec{d} = \tens{T} \vec{s} + \vec{n},
\end{equation}
where $\vec{s}$ is the true sky signal projected from the sky map as $\vec{s} = \tens{P}\vec{m_{\vec{s}}}$; $\tens{P}$ is a pointing matrix of size $(n_{\mathrm{pix}}, n_{\mathrm{tod}})$ that maps between pixel space and time ordered data; $\tens{T}$ represents convolution with a bolometer transfer function $\tens{T} = \tens{F}^{-1}\tens{T}(\omega)\tens{F}$, where $\tens{F}$ denotes a Fourier transform; and $\vec{n}$ denotes noise. In this work, we assume that the latter only consists of zero-mean Gaussian noise, and we define its covariance matrix as $\tens{N}\equiv\left<\vec{n}\vec{n}^T\right>$. However,  there are many other sources of instrumental noise in the real data (e.g., \citealt{planck2014-a08}, \citealt{BP_VI}) that must be taken into account in a full analysis pipeline. 

Our goal now is to derive an accurate estimate of $\vec{s}$ given some observed data, and we will denote this estimate $\hat{\vec{m}}$. The traditional approach for doing this adopted by most bolometer-based experiments consists of a two-step procedure; one specific example that is particularly relevant for this paper is \Planck\ HFI \cite{planck2013-p03c, planck2014-a08}. The first step is to explicitly apply the inverse transfer function operator to the raw TOD,
\begin{equation}
    \tens{T}^{-1} \vec{d} = \tens{T}^{-1}(\tens{T} \vec{s} + \vec{n}) = \vec{s} + \tens{T}^{-1} \vec{n}.
    \label{eq:deconvolved_tod}
\end{equation}
As long as $\tens{T}$ is non-singular, this results in an unbiased estimate of the signal $\vec{s}$. However, it also modifies the noise, $\vec{n}$. In particular, due to the shape of the transfer function (as shown in Fig.~\ref{fig:TF_vs_filter}), this inverse operator significantly boosts the noise level at high frequencies in Fourier space. To prevent the introduction of excessive noise, a common solution is to apply an additional low-pass filter function $K(\omega)$, such that the total filtered TOD reads
\begin{equation}
    \tens{K}\tens{T}^{-1} \vec{d} = \tens{K}\tens{T}^{-1}(\tens{T} \vec{s} + \vec{n}) = \tens{K}\vec{s} + \tens{K}\tens{T}^{-1} \vec{n}.
    \label{eq:deconvolved_tod_with_filter}
\end{equation}
Here we have introduced a filter operator $\tens{K}$ similar to the transfer function operator $\tens{T}$ as $\tens{K} = \tens{F}^{-1}\tens{K}(\omega)\tens{F}$. 

The second step in the traditional procedure is to apply a mapmaking algorithm to this deconvolved TOD. Under the assumption of Gaussian noise, the optimal solution for this is given by the normal equations \citep[e.g.,][]{tegmark_1997},
\begin{equation}
    \label{eq:classic_mapmaking}
    \tens{P}^{T} \tens{N}^{-1} \tens{P} \hat{\vec{m}}_{\mathrm{trad}} = \tens{P}^{T} \tens{N} ^{-1}\vec{d}.
\end{equation}
In practice, this optimal mapmaking equation is often replaced with a computationally cheaper solution that does not require inversion of a full dense noise covariance matrix. In many cases $\mathrm{N}$ is simply approximated with its diagonal, and in that case the equation may be solved pixel-by-pixel (so-called ``binning''). Another common solution is to apply a destriping algorithm, which accounts for large-scale noise fluctuations. In either case, $\hat{\vec{m}}_{\mathrm{trad}}$ is sub-optimal in two respects: First, if $\tens{K}\ne\tens{I}$, then $\hat{\vec{m}}_{\mathrm{trad}}$ is a biased estimator of $\vec{s}$. In practice, this is typically accounted for in higher-level analysis by modifying the effective instrumental beam, which then introduces significant asymmetries that couple to the scanning strategy. Second, the actual noise covariance matrix in the post-deconvolved TOD reads $\tens{K}\tens{T}^{-1}\tens{N}\tens{T}^{-1}\tens{K}$, but these additional terms are not accounted for in the above solution. As such, the noise weighting of $\hat{\vec{m}}_{\mathrm{trad}}$ is also sub-optimal.

Aiming to resolve both these deficiencies, we adopt a simpler approach in this paper, and note that an unbiased and optimal estimate of $\hat{\vec{m}}$ can be obtained directly from the data model in Eq.~\eqref{eq:TOD} as follows,
\begin{equation}
    \label{eq:mapmaking}
    \tens{P}^{T} 
    \tens{T}^{T}
    \tens{N}^{-1} 
    \tens{T}
    \tens{P} \hat{\vec{m}} = \tens{P}^{T}
    \tens{T}^{T}
    \tens{N}^{-1} \vec{d}.
\end{equation}
We denote the solution of this equation $\hat{\vec{m}}_{\mathrm{MLE}}$, where MLE is short of ``maximum likelihood estimate''. The equation involves the transpose of $\tens{T}$, which may be written as $\tens{T}^{T} = \tens{F}^{-1}\tens{T}^{*}(\omega)\tens{F}$, where  $\tens{T}^{*}(\omega) = T^{*}(\omega)$ is the complex conjugate of the transfer function $T(\omega)$.

We use a standard preconditioned conjugate gradient method
(CG; \citealp{CG_pain}) to solve Eq.~\eqref{eq:mapmaking}, and we find
that a simple diagonal preconditioner of the form
\begin{equation}
    \label{eq:preconditioner}
    \tens{M}=\tens{P}^T \tens{P} = \sum_{tt'} \tens{P}_{tp} \tens{P}_{t'p}
\end{equation}
results in a speed-up of almost a factor of 10 compared to no preconditioning. This algorithm is conceptually identical to the \texttt{artDeco} algorithm introduced by \citet{ArtDeco} for asymmetric beam deconvolution, the main difference being that our $\tens{T}$ operator is computationally much cheaper than their asymmetric beam operator $\tens{B}$.

\section{One-dimensional toy model: Intuition}
\label{sec:1D_test}

\begin{figure}
    \centering
    \includegraphics[width=\linewidth]{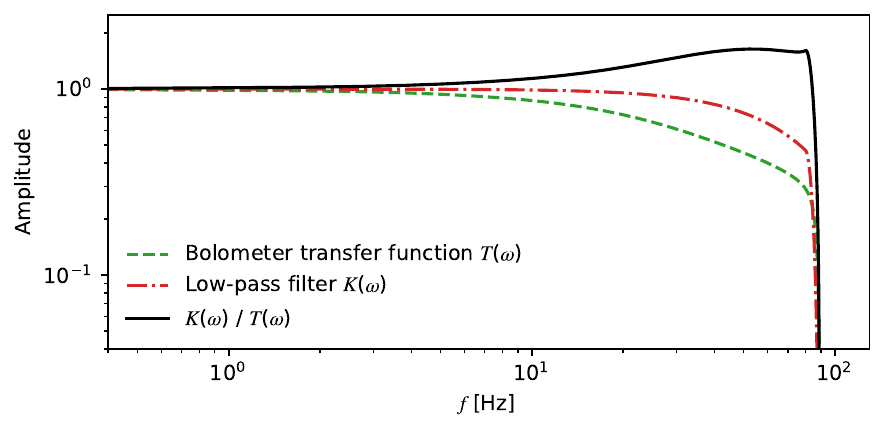}
    \caption{Amplitudes of the \Planck\ HFI bolometer transfer function $T(\omega)$ (green), the low-pass filter $K(\omega)$ employed by \Planck\ HFI to keep the high-k modes from blowing up (red), and the former divided by the latter (black), which is the resulting function applied to the data in \Planck\ deconvolution method. The presented functions are in absolute amplitudes. The bolometer transfer function also has a complex component.}
    \label{fig:TF_vs_filter}
\end{figure}

In general, our own primary motivation for this line of work lies in a future reanalysis of the \Planck\ HFI observations. We therefore adopt the HFI 143-5 bolometer transfer function $T(\omega)$ and low-pass filter $K(\omega)$ as an explicit test case, which is shown in Fig.~\ref{fig:TF_vs_filter}. The goal of this and the following two sections is to compare the performance of the traditional and the optimal methods in various settings for this case. In fact, most of the key algebraic points can be easily demonstrated and visualized through a simple one-dimensional case in which all matrix operations can be solved quickly by brute-force methods

In this first example, we define our true input sky map to consist of an array with 200 one-dimensional pixels. This signal map is then scanned by a simple sinusoidal scanning strategy, and the resulting signal-only TOD is convolved with the \Planck\ transfer function shown in Fig.~\ref{fig:TF_vs_filter}. Finally, white Gaussian noise is added. We then solve Eqs.~\eqref{eq:classic_mapmaking} and \eqref{eq:mapmaking} for $\hat{\vec{m}}_{\mathrm{trad}}$ and $\hat{\vec{m}}_{\mathrm{MLE}}$, respectively. Since the number of pixels is only 200, these solutions are very fast even with brute-force matrix inversion.

\begin{figure}
    \centering
    \includegraphics[width=\linewidth]{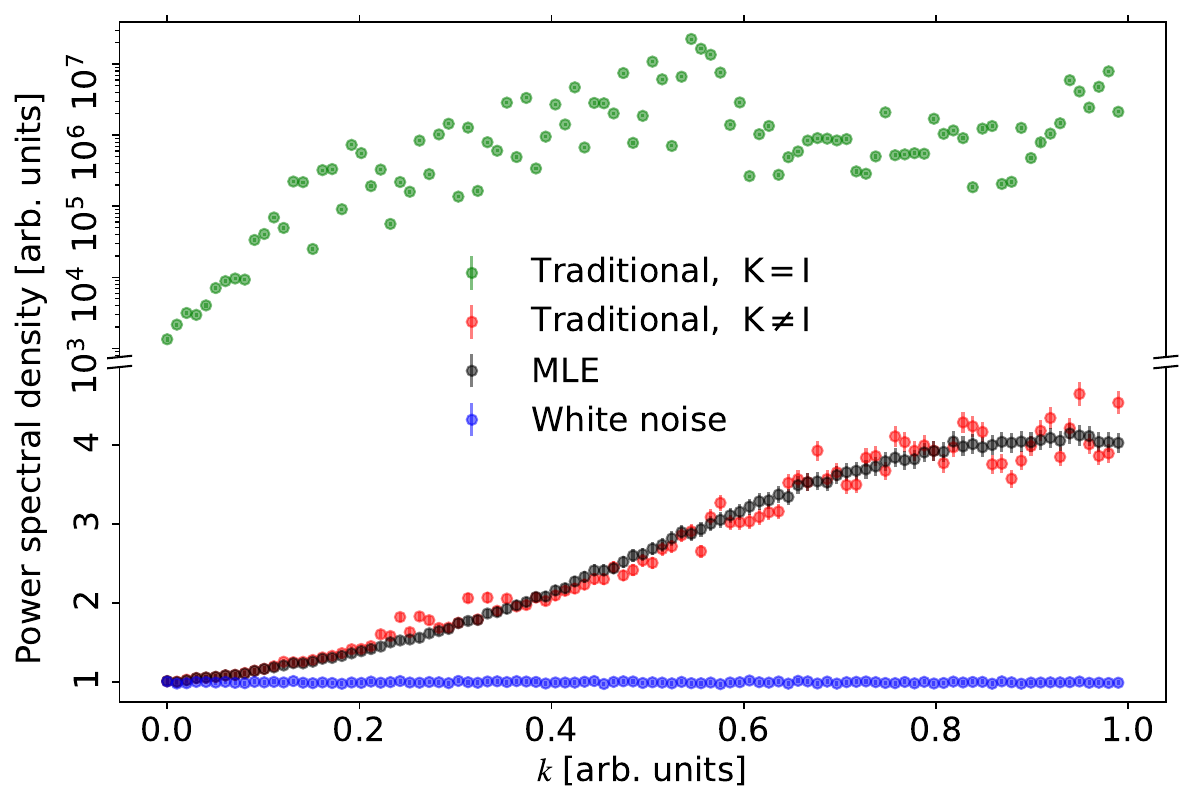}
    \caption{Comparison of noise power spectra for noise-only 1D simulations for four different analysis configurations, as evaluated by the mean and standard deviation from 10\,000 simulations. The $y$-axis is broken into linear and logarithmic portions.}
    \label{fig:1D_toy_PS_results}
\end{figure}

\begin{figure}
    \centering
    \includegraphics[width=\linewidth]{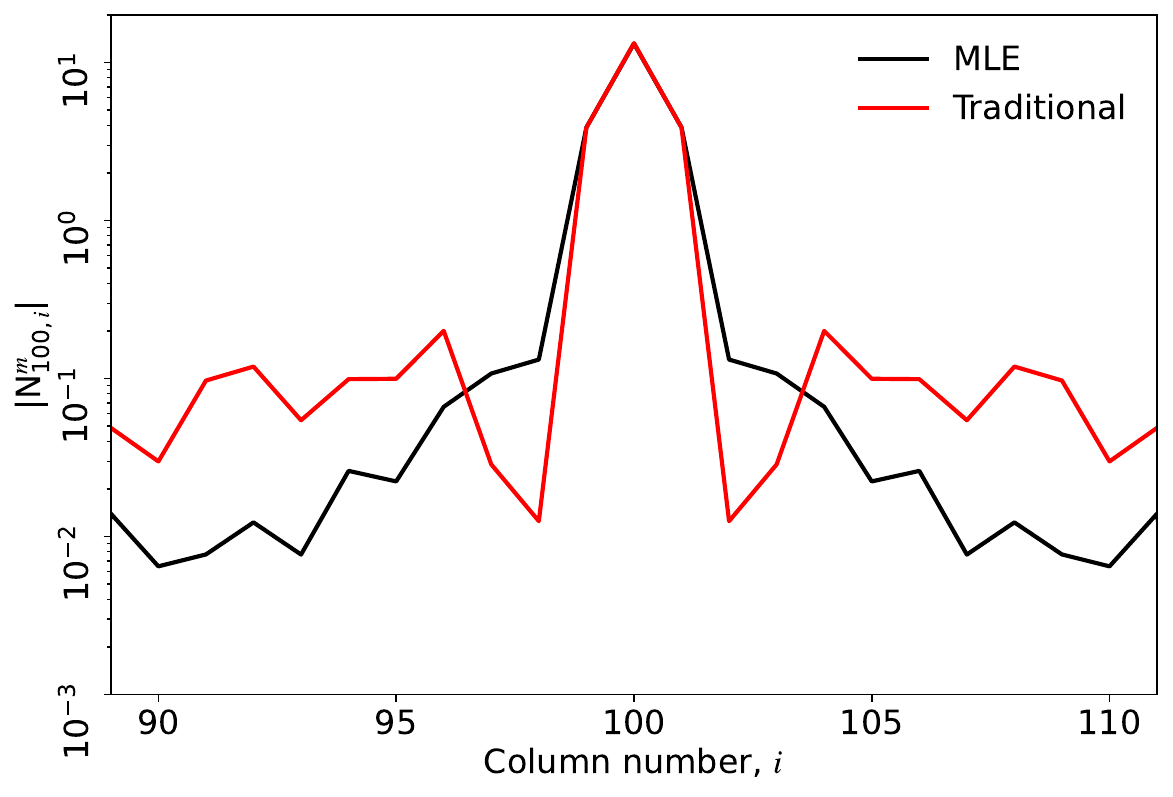}
    \caption{Absolute value of a middle row map covariance matrix slice, $|\tens{N}^{m}_{100, i}| = |(\tens{P}^T \tens{N}^{-1} \tens{P})^{-1}_{100, i}|$, for both the 1D MLE (black) and binned and low-pass filtered solutions of the 1D toy model (red). These covariances have been found from the corresponding map-domain power spectra seen in Fig. \ref{fig:1D_toy_PS_results}.}
    \label{fig:1D_toy_PS_results_covariance}
\end{figure}

We consider two different cases for this 1D model. In the first case, we set $\vec{s}=0$, with the goal of understanding the effects of the different deconvolution methods on the noise properties of the resulting maps. In the second case, we insert a single narrow Gaussian peak in the middle of the map, representing a point source in a typical CMB map, with the goal of understanding the relative impact of the two methods on a sharp signal.




\begin{figure}
    \centering
    \includegraphics[width=\linewidth]{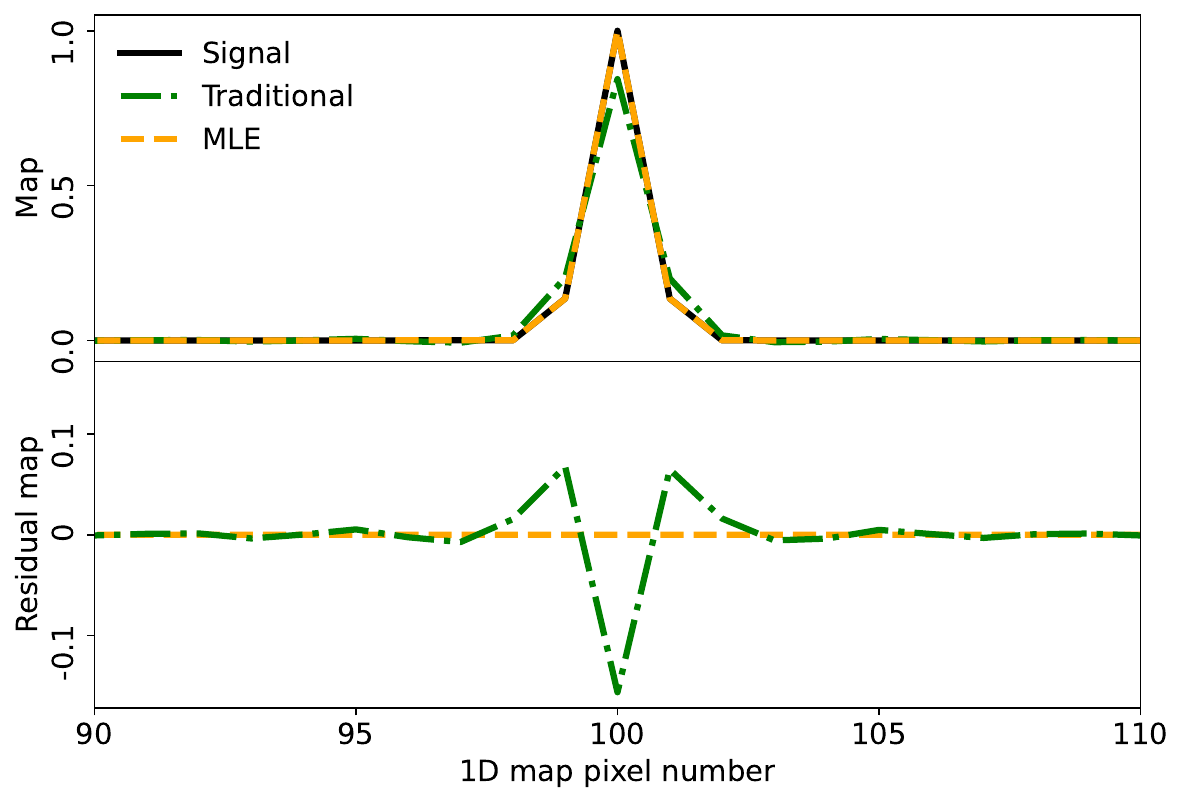}
    \caption{Comparison of reconstructed 1D point source signals for $\hat{\vec{m}}_{\mathrm{trad}}$ (dashed green) and $\hat{\vec{m}}_{\mathrm{MLE}}$. The top panel shows the full reconstructed signal amplitude, with the true input shown as a solid black line, and the bottom panel shows the difference between output and input signals.}
    \label{fig:1D_point_source_map}
\end{figure}

Starting with the noise-only case, we simulate a total of 10\,000 independent noise realizations, and use these to build up the post-solution noise covariance matrix explicitly. For each realization, we calculate the power spectrum defined as $P(k) = \langle|f(k)|^2\rangle$, where $f(k)$ is the Fourier transform of $\hat{\vec{m}}$. In the case of $\hat{\vec{m}}_{\mathrm{trad}}$, we have to deconvolve the effective transfer function arising from $\tens{K}$ to obtain an unbiased estimate. This is found by simulating a large ensemble of random signal-only maps and taking the ensemble average of the ratio between the corresponding output and input spectra. To illustrate the adverse impact of unregularized high-frequency noise on $\hat{\vec{m}}_{\mathrm{trad}}$, we also include a case corresponding to $\tens{K}=\tens{I}$ for $\hat{\vec{m}}_{\mathrm{trad}}$ in this demonstration.

The results from these calculations are summarized in Fig.~\ref{fig:1D_toy_PS_results}. For reference, the blue curve shows the power spectrum of a white noise TOD directly binned into a map, without taking into account $\tens{T}$; this illustrates the intrinsic noise level that is subsequently boosted by the bolometer transfer function deconvolution in the actual methods. Starting from the top, the green dots shows the noise in $\hat{\vec{m}}_{\mathrm{trad}}$ when not applying the regularization kernel $\tens{K}$. The $y$-axis in the plot is broken into linear and logarithmic scaling. The high-frequency noise must be suppressed prior to mapmaking in some way or other to obtain meaningful results. Moving onto the realistic cases corresponding to $\hat{\vec{m}}_{\mathrm{trad}}$ with $\tens{K}$ and $\hat{\vec{m}}_{\mathrm{MLE}}$ shown in red and black, respectively, we see that the two methods perform similarly in terms of total noise power. 

However, even though the two methods perform similar in terms of absolute noise power, they still perform quite differently in terms of noise correlations. This is illustrated in Fig.~\ref{fig:1D_toy_PS_results_covariance}, which shows a slice through the empirical correlation matrix evaluated as $\left<\hat{\vec{m}}_j\hat{\vec{m}}_j^T\right>$ over the simulated ensemble, where $j$ indicate simulation number. We see that the correlation falls off by almost an order of magnitude lower values for $\hat{\vec{m}}_{\mathrm{MLE}}$ than $\hat{\vec{m}}_{\mathrm{trad}}$ at long distances; the low absolute values for $\hat{\vec{m}}_{\mathrm{trad}}$ at a few pixels separation is just a ringing artifact from the $\tens{K}$ filter. 


Next, we consider the signal case with a Gaussian point source in the middle of the 1D map. We repeat the same procedure as outlined in the previous section for both techniques. However, since we are now interested in the effect on the signal, and the mapmaking equations are linear, we now omit the noise in the actual simulated TOD. The resulting products therefore correspond directly to ensemble-averaged quantities, and require no Monte Carlo simulation. The outputs from these calculations are summarized in Fig.~\ref{fig:1D_point_source_map}. The top panel shows the input model as a solid black curve, and the reconstructed estimates are shown as dashed orange (for $\hat{\vec{m}}_{\mathrm{MLE}}$) and dashed green (for $\hat{\vec{m}}_{\mathrm{trad}}$) curves. The bottom panel shows the difference between output and input signals. The input signal is normalized to unity at the peak, so that the bottom panel can be interpreted as a fractional error. The MLE solution results in an unbiased estimate of the input signal, and any uncertainties in this solution are given by numerical round-off errors. In contrast, the traditional method results in residuals at the 10\,\% level at the peak, with significant ringing extending to large distances. 


\section{Two-dimensional toy model: Impact on effective beam}
\label{sec:point_source}

In this section we apply the methods outlined in Sect.~\ref{sec:method} to a two-dimensional case, with the goal of comparing the performance of the traditional and the MLE methods in terms of their impact on the effective beam. In this case, we construct an input sky signal that is mostly empty, except for one or more bright point sources depending on the test in question. In the first case we study a single point source located at the Ecliptic South Pole, and the input map is defined by setting the closest pixel to 100 in arbitrary units. The map is then smoothed with a symmetric Gaussian beam with the full width at half maximum (FWHM) set to $7.2$ arcmin, similar to the 143\,GHz \textit{Planck} HFI effective beam \citep{planck2013-p03c}. 

\begin{figure}
    \centering
    \includegraphics[width=\linewidth]{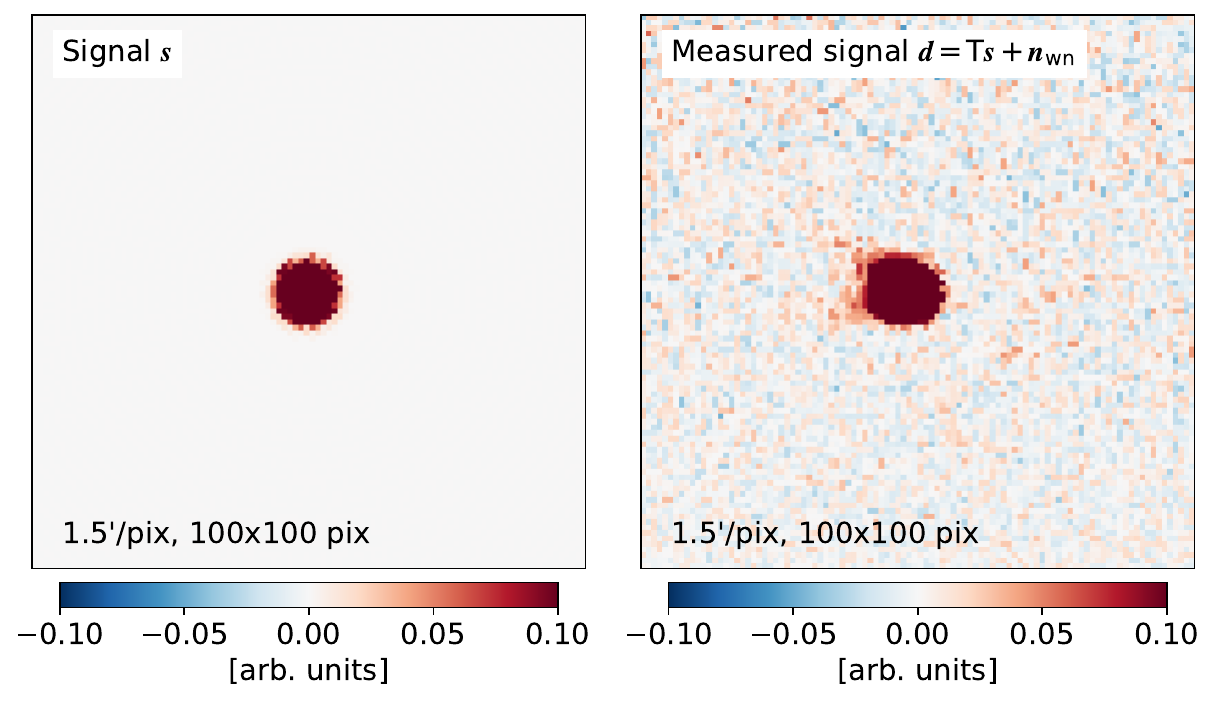}
    \caption{Simulated point source signal $\vec{s}$ on the \emph{left} panel and detector measured signal $\vec{d}=\tens{T} \vec{s} + \vec{n}$ on the \emph{right} panel. The effect of the transfer function $\tens{T}$ can be observed in the smearing of the point source signal along the scanning path, resulting in a deformed beam.}
    \label{fig:point_source}
\end{figure}

Using this map, the sky signal TOD is created using the \textit{Planck} 143-5 pointing matrix as $\vec{s} = \tens{P} \vec{m}_{\rm{sky}}$. For the purposes of this experiment, which is designed to build intuition regarding the effect of a bolometer transfer function on a point source, we only use the first three months of the HFI survey. The measured TOD $\vec{d}$ is created through Eq.~\eqref{eq:TOD} by applying the bolometer transfer function and adding white noise. The white noise level is set to $\sigma_{\rm{wn}} = 0.3125$ in arbitrary units, which corresponds to a signal-to-noise ratio of 320 at the peak of the point source. This value is chosen to represent the typical signal-to-noise ratio of planet observations reported by \citet{planck2013-p03c}. The maps for both the input sky signal $\vec{s}$ and the corresponding naively binned sky map $\hat{\vec{m}}_{\mathrm{bin}}$ are shown in Fig.~\ref{fig:point_source}. In the first case, the point source appears azimuthally symmetric, while in the second case it is significantly deformed. Because the satellite scanning moved from right to left in this figure,  the transfer function effectively drags the signal along the scanning path. 

\begin{figure}
    \centering
    \includegraphics[width=\columnwidth]{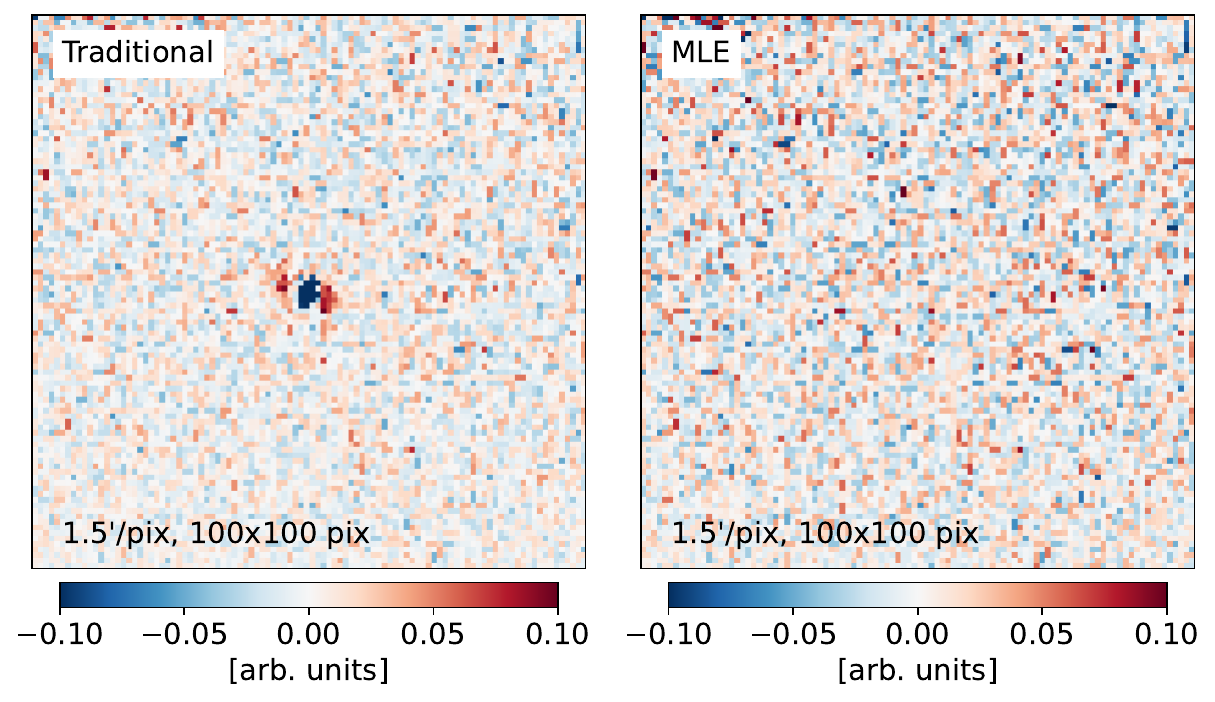}
    \caption{Residual comparison between the traditional deconvolution method and MLE based deconvolution method. The \emph{left} panel shows the residual between the signal deconvolved using the inverse of the transfer function operator $\tens{T}^{-1}$ in combination with filter $K(\omega)$ and the original point source signal $\vec{s}$ (\emph{left} panel in Fig. $\ref{fig:point_source}$). The \emph{right} panel shows the residual between the signal deconvolved by solving Eq. $\eqref{eq:mapmaking}$ and the original point source signal.}
    \label{fig:point_source_comp}
\end{figure}

We now apply both the traditional and the MLE methods to these simulated data, each producing a map of the true sky signal. The results from these calculations are shown in Fig.~$\ref{fig:point_source_comp}$ in terms of the difference between the reconstructed and the true input maps. Starting with the traditional method, we observe at least two noticeable residuals related to the transfer function. First, since the traditional method includes a regularization kernel $\tens{K}$, the algorithm is unable to reconstruct the true input sky signal, and a quadrupolar residual aligned with the scanning path is present in the residual map. For an isotropic and random field, such as the CMB, the average effect of this can be accounted for by modifying the effective azimuthally symmetric beam response function $b_{\ell}$, as for instance is done in the \Planck\ analysis, but this method is clearly unable to reconstruct an optimal image of the true sky. In contrast, the integrated MLE solution shows no signs of scan-aligned effects, and the residual is consistent with white noise. Secondly, far away from the point source, the traditional method smooths the white noise more than the MLE method. Both of these effects correspond directly to what was found for the one-dimensional toy model in the previous section.


The magnitude of this effect depends on the detailed scanning path of the instrument, and hence with the position on the sky. We now aim to quantify the effective beam deformation produced by the bolometer transfer function for both methods in terms of the effective beam FWHM and ellipticity over the full sky. For these purposes a statistically meaningful number of point sources is required. Therefore, we  create a map with 12\,288 point sources, each located at the center of a HEALPix $N_{\mathrm{side}}=32$ map. The analysis itself is performed with $N_{\mathrm{side}}=2048$, corresponding to a pixel size of $1\farcm7$. Then we apply exactly the same process as earlier in this section in terms of transfer function operator $\tens{T}$, white noise, and map-making methods.


For each point source and both methods, we measure the effective FWHM and ellipticity by fitting a two-dimensional Gaussian following the steps similar to \cite{Fosalba_2002}. We define the ellipticity parameter $\varepsilon = \sigma_{\mathrm{long}}/ \sigma_{\mathrm{short}}$ as the ratio between the long and short axes of the ellipse. An azimuthally symmetric object corresponds to $\varepsilon = 1$, while $\varepsilon > 1$ corresponds to a deformed beam. In terms of polar coordinates $(\rho , \phi)$, the actual function fitted to each two-dimensional object is
\begin{equation}
    \label{eq:ellipticity_fit}
    z(\rho, \phi) = A \cdot \exp \left[- \frac{\rho^{2}}{2\sigma_{\mathrm{short}}^{2}} \left(1 - \chi \cdot \cos (\phi - \alpha)^{2}\right) \right],
\end{equation}
where $\sigma_{\mathrm{short}}$ is the width of the short axis of the ellipse, $\chi \equiv 1 - 1/\varepsilon^{2}$, and $\alpha$ is the rotation angle to align coordinate axes with the ellipse axes. The effective FWHM is defined as $\sqrt{8\ln 2}$ times the average between the long and short axes widths, where long axis width can be found from the relation for ellipticity $\sigma_{\mathrm{long}} = \varepsilon \cdot \sigma_{\mathrm{short}}$. 

The function defined in Eq.~\eqref{eq:ellipticity_fit} assumes that the coordinate center is located at the peak, and this is not necessarily true after beam convolution. We therefore run an additional fit for the center pixel coordinates from the $N_{\mathrm{side}}=32$ map to the center coordinates of these beams on the $N_{\mathrm{side}}=2048$ map. We can assume local space around a given beam to be Euclidean. Then Eq.~\eqref{eq:ellipticity_fit} follows from a regular two-dimensional normal distribution in Cartesian coordinates after the polar coordinate transformation. In order to align the axes of the ellipse with the smearing effect produced by the deconvolution, a rotational angle $\alpha$ is introduced into the fitting function.

\begin{figure}
    \centering
    \includegraphics[width=\columnwidth]{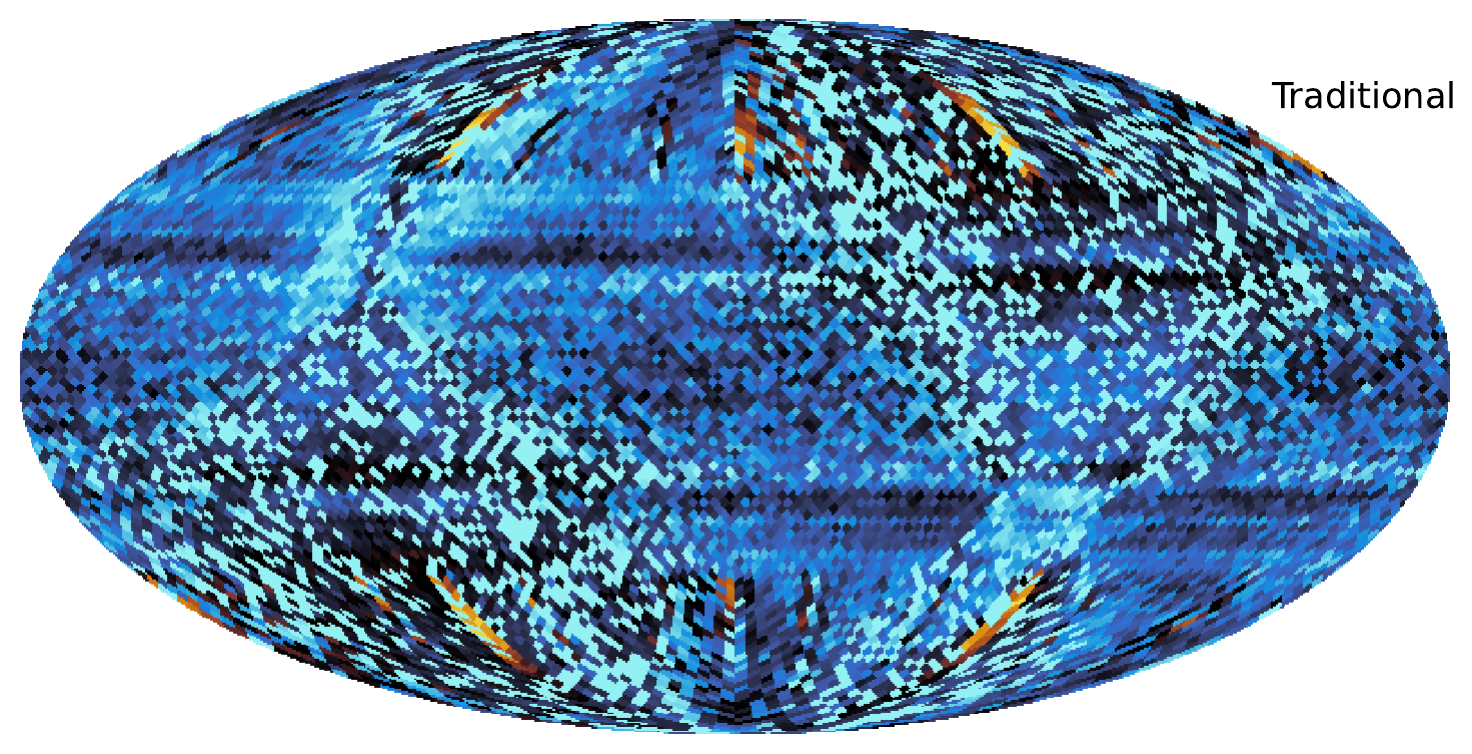}
    \includegraphics[width=\columnwidth]{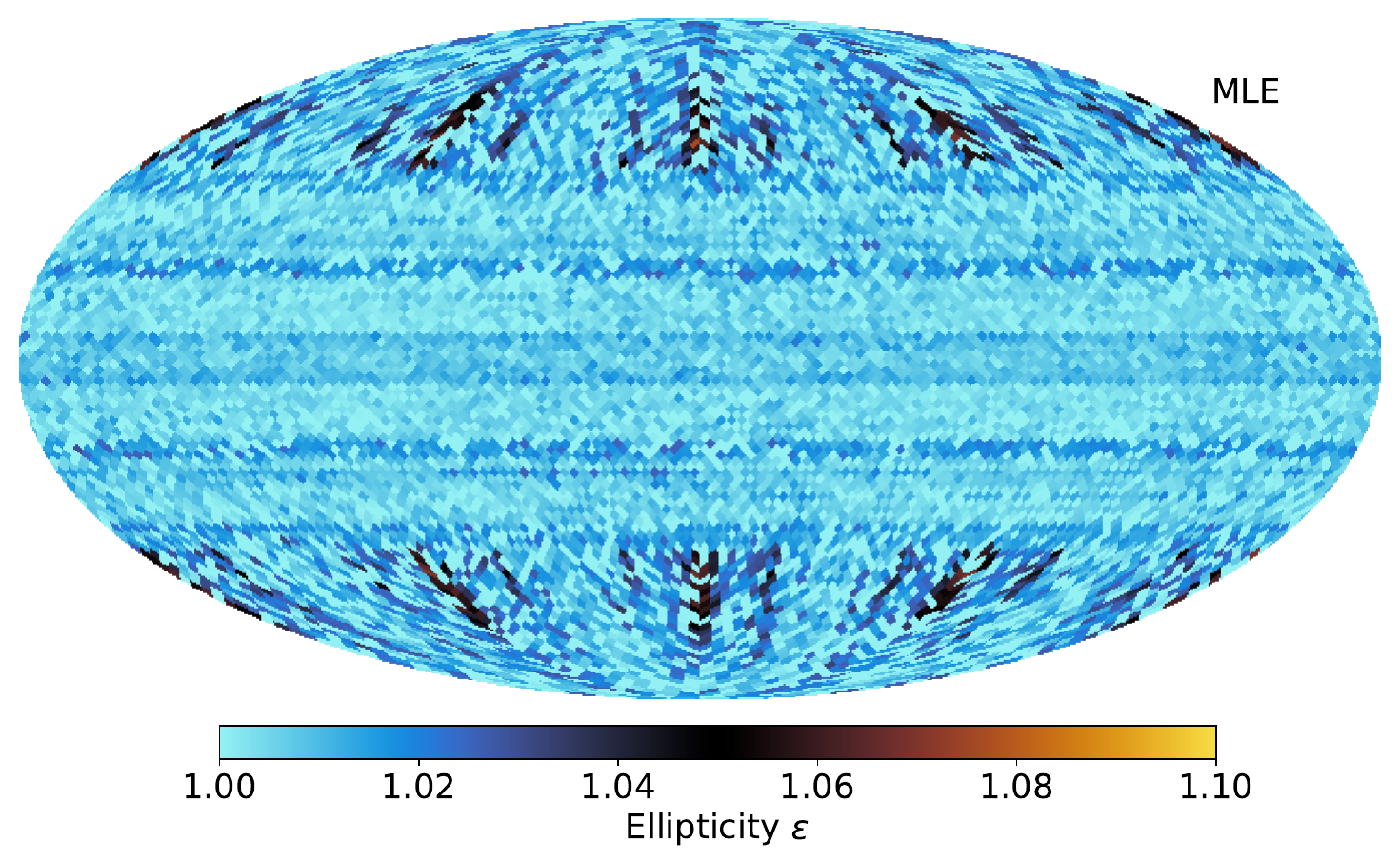}
    \caption{Distribution of ellipticity  $\varepsilon = \sigma_{\mathrm{long}}/ \sigma_{\mathrm{short}}$ over the sky. The ellipticity was measured as a parameter in a Gaussian fit in Eq. $\eqref{eq:ellipticity_fit}$. The \emph{upper} panel shows the ellipticity of the beams deconvolved using traditional method, and the \emph{lower} panel - deconvolved with MLE method.}
    \label{fig:ellipticity_maps}
\end{figure}

In Fig.~\ref{fig:ellipticity_maps} we show the distribution of $\varepsilon$ over the full sky for both mapmaking methods, and we notice that the traditional method results in a noticeably higher ellipticity across the sky compared to the MLE method proposed in this paper, and it has a much stronger imprint of the \Planck\ scanning strategy. In contrast, the distribution seen for the MLE method is defined primarily by the underlying HEALPix grid, which is unavoidable given the choice of pixelization. 

\begin{figure}
    \centering
    \includegraphics[width=\columnwidth]{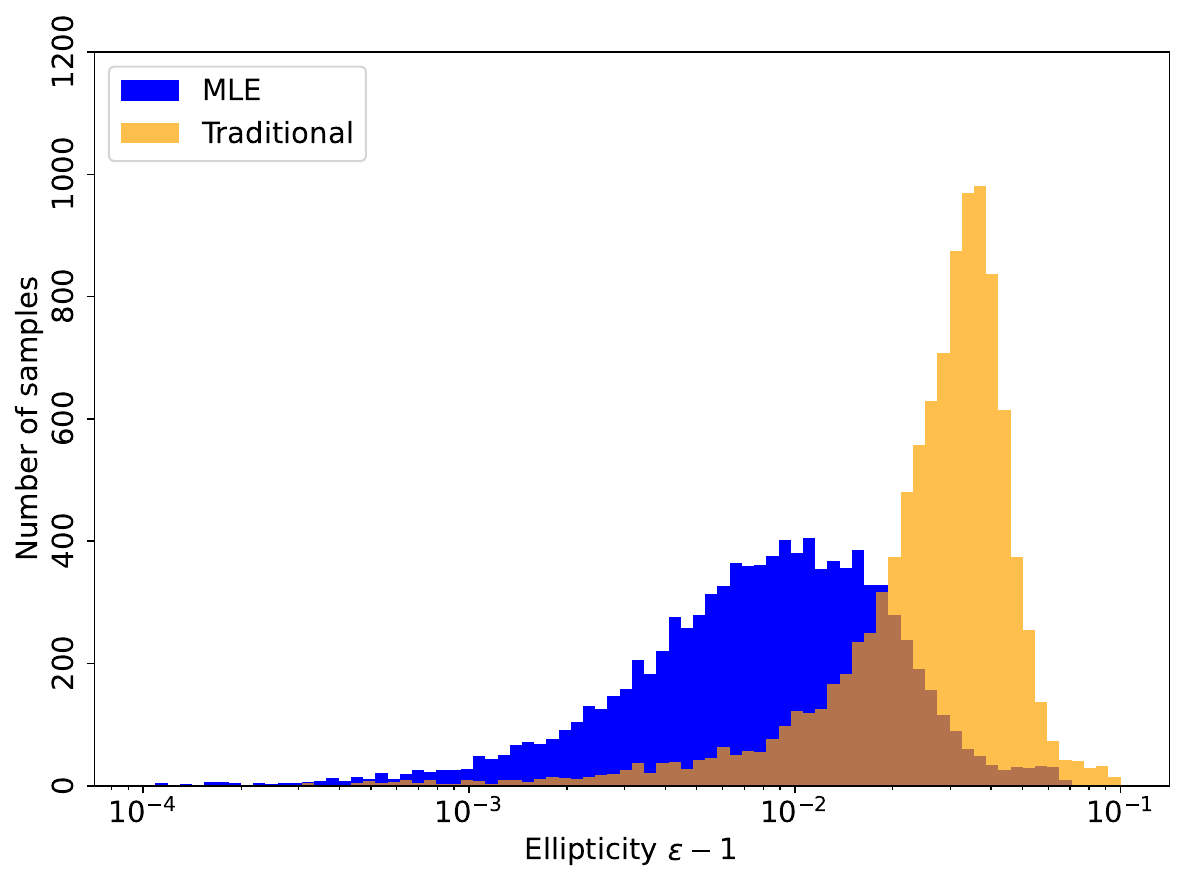}
    \caption{Distribution of effective ellipticities, $\varepsilon-1$, for the traditional (orange histogram) and MLE mapmaking (blue histogram) algorithms.}
    \label{fig:ellipticity_hist}
\end{figure}

Figure ~\ref{fig:ellipticity_hist} shows the same information in terms of histograms of $\varepsilon-1$. The corresponding means and standard deviations for the two distributions are $\varepsilon_\mathrm{trad}-1 = 0.025 \pm 0.014$ and $\varepsilon_{\mathrm{MLE}}-1 = 0.009 \pm 0.010$. The mean ellipticity of the MLE method is thus 65\,\% smaller than for the traditional method. The values found for the traditional method are close to those reported by \cite{planck2013-p03c} for the 143\,GHz channel.

\begin{figure}
    \centering
    \includegraphics[width=\columnwidth]{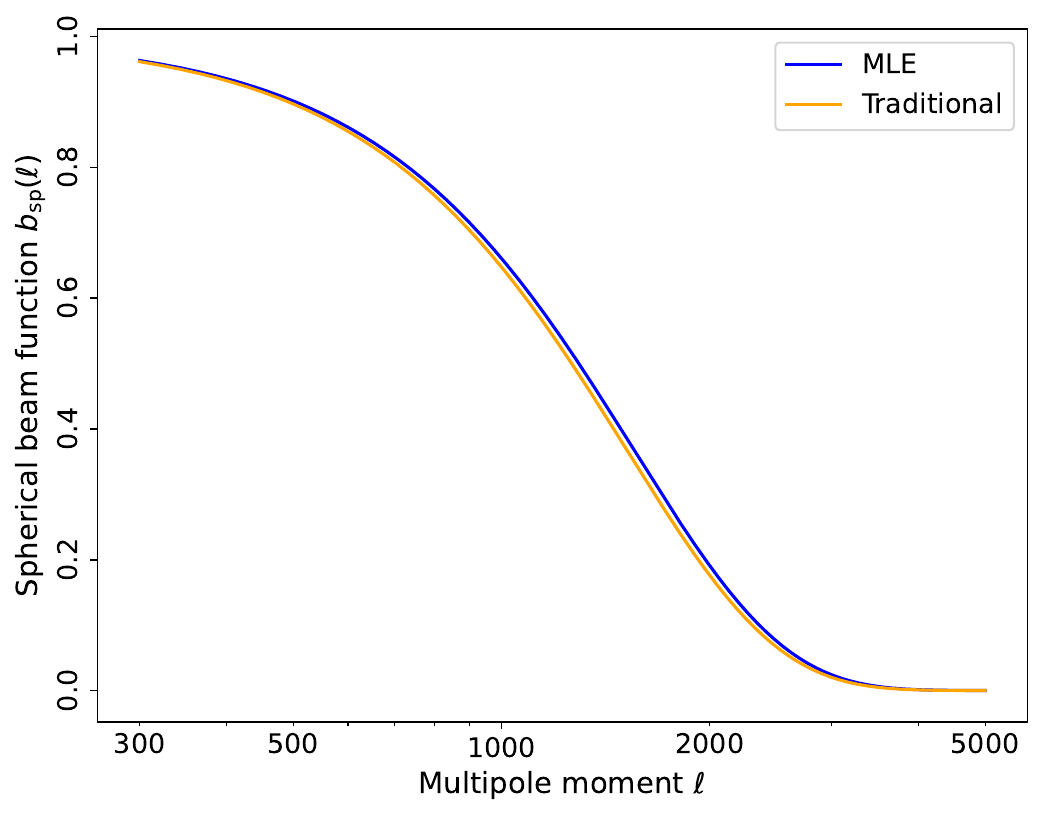}
    \caption{Spherical beam function for two deconvolution methods. These functions are calculated based on the average FWHM for each method. The blue line shows the spherical beam function for the maximum likelihood deconvolution method, while the  orange line shows \Planck\ method.}
    \label{fig:FWHM}
\end{figure}

Performing a similar comparison for the effective FWHM, we find that the MLE method results in a 2.3\,\% lower value than the traditional method. The net impact of this difference in terms of effective beam transfer functions, $b_{\ell}$, is shown in Fig.~\ref{fig:FWHM}, where
\begin{equation}
    b_{\mathrm{sp}}(\ell) = \exp \left( -\frac{\ell (\ell+1) \mathrm{FWHM}}{16 \ln 2}\right).
\end{equation}
At $\ell=2500$, the ratio between these two functions is 1.14, while at $\ell=4000$ it is 1.38.  


\section{CMB simulation}
\label{sec:CMB_case}

Finally, we consider a semi-realistic CMB-plus-noise case. In this case, the sky signal $\vec{s}$ is generated as a Gaussian random realization based on a best-fit $\Lambda$CDM temperature power spectrum computed with CAMB \citep{Lewis_1999, Howlett_2012} and adopting best-fit parameters from \citet{planck2016-l05}. The simulated TOD is then generated by observing this map with the full-mission \Planck\ 143-5 scanning path, and applying the corresponding bolometer transfer function $\tens{T}$. Finally, we add white noise with $\sigma = 200$ $\mu$K per sample. This value corresponds to co-adding all 143\,GHz bolometers into one, such that our final simulation has similar sensitivity as the true 143\,GHz frequency map, but the data volume of only a single detector. To allow for the calculation of cross-power spectra, we split the data into two halves, and process each half independently.

\begin{figure}
    \centering
    \includegraphics[width=\columnwidth]{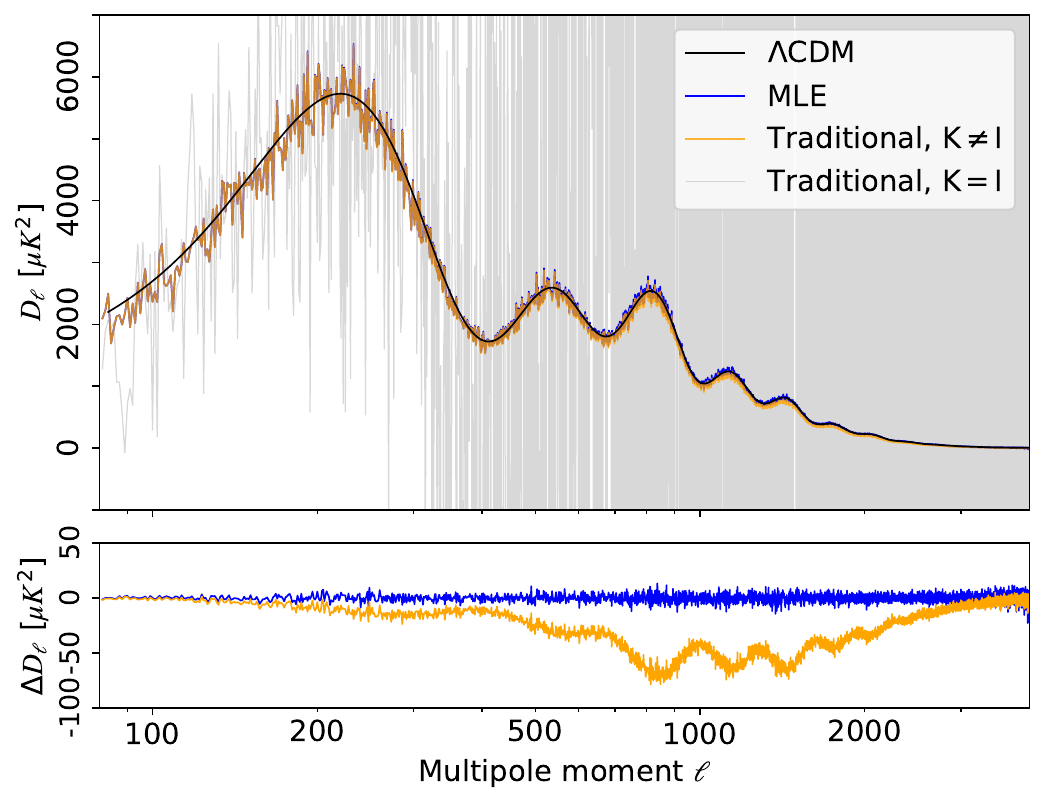}
    \caption{(\emph{Upper} panel) Power spectra $D_{\ell}$ of the deconvolved CMB maps. The original signal was simulated based on the $\Lambda$CDM spectrum shown with the black solid line. The blue line shows the power spectrum obtained from the maximum likelihood based deconvolution. The orange line shows the power spectrum of the map obtained from the traditional deconvolution process $\tens{K}\tens{T}^{-1}\vec{d}$. The grey line shows the power spectrum of the map obtained from the unfiltered ($\tens{K} = \tens{I}$) traditional deconvolution process $\tens{T}^{-1}\vec{d}$. All spectra are calculated as cross power spectra using two noise realisation $\vec{n}^{1,2}_{\rm{wn}}$, to exclude the white noise power spectrum.
    (\emph{Lower} panel) Difference between the deconvolved power spectra and input $\Lambda$CDM power spectrum. The colors represent the difference for the respective deconvolution method in the \emph{upper} panel.}
    \label{fig:power_spectra_CMB}
\end{figure}

We now apply the same three map-making methods to this TOD simulation as shown in Fig.~\ref{fig:1D_toy_PS_results} for the one-dimensional case, namely the traditional method (with and without a regularization kernel) and the new MLE method. The results from this calculation are summarized in Fig.~\ref{fig:power_spectra_CMB} in terms of cross-angular power spectra $D_{\ell} = C_{\ell}\ell(\ell+1)/2\pi$.

The traditional method without a low-pass filter $\tens{K}$ (gray) produces results that are extremely noisy even after taking the cross-power spectrum, mirroring the 1D case shown in Fig.~\ref{fig:1D_toy_PS_results}. Secondly, we see that both the traditional and MLE methods reproduce the original $\Lambda$CDM power spectrum at low multipoles, $\ell\lesssim 100$. However, at higher multipoles the traditional method is noticeably lower. This deviation is caused by the additional $\tens{K}$ smoothing operator, which has not been deconvolved in this plot. 



\begin{figure}
    \centering
    \includegraphics[width=\columnwidth]{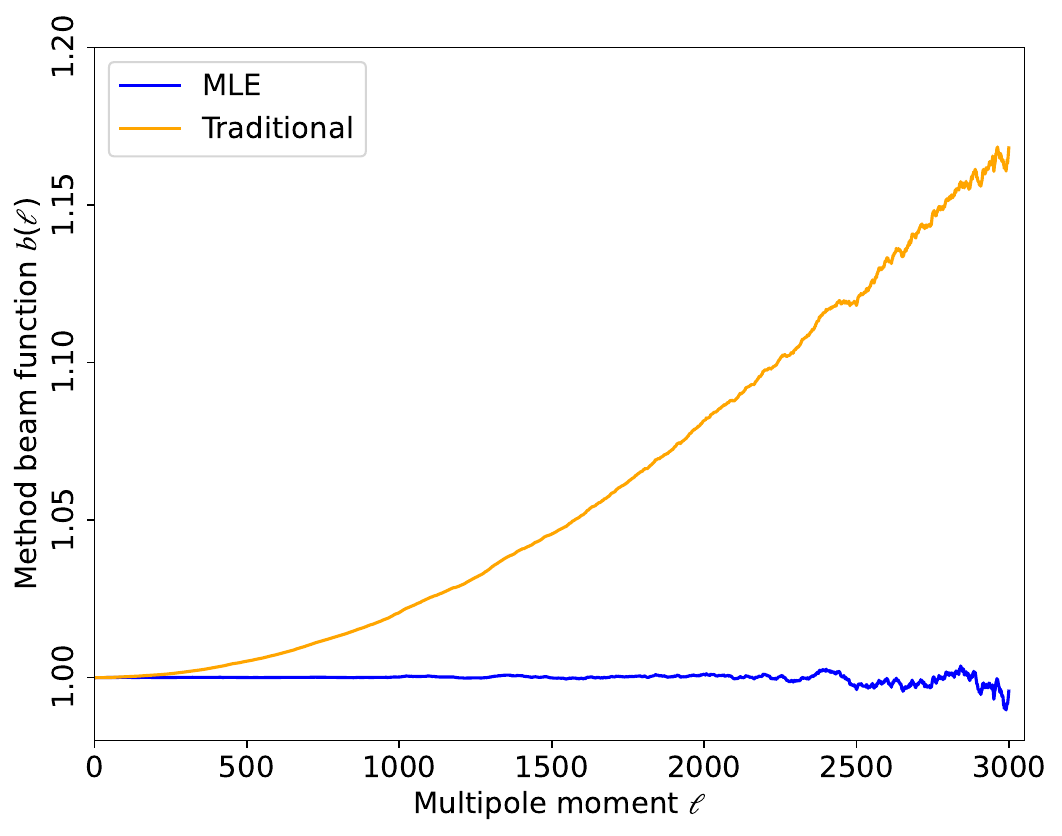}
    \caption{Effective beam function introduced by the deconvolution method. This function is found as the square of a ratio between the simulated $\Lambda$CDM power spectrum and the power spectrum calculated for the respective deconvolution method. The maximum likelihood based method produces a beam function equal to one, as shown by the blue line. The traditional deconvolution method results in the beam function increasing with the multipole moment $\ell$ and is illustrated with the orange line.}
    \label{fig:beam_func}
\end{figure}

In order to correct for this bias, one has to include the effect of $\tens{K}$ into the effective beam profile, as for instance done by \citet{planck2013-p03c}. To measure the total beam profile, we simply calculate the square root of the ratio between output and input power spectra. The results are shown in Fig.~\ref{fig:beam_func}. Here we see that the MLE method produces an effective beam profile that is very close to unity for almost all multipoles. The small deviations seen at higher $\ell$ are due to instrumental and numerical noise. On the other hand, the traditional method results in a ratio that monotonically increases with $\ell$. 


\section{Computational expense}
\label{sec:cost}


The code used in this this paper was written in Python, and relies on utilities provided by the \texttt{scipy} \citep{Scipy} and \texttt{numpy}  \citep{numpy} libraries. The majority of the runtime is spent on Fast Fourier Transforms (FFT), which are performed with a compiled C++ code under the hood. However, the emphasis in this paper has been the fundamental algebraic solution, and not code optimization, and the runtimes quoted in the following can very likely be improved by a significant factor in a future production implementation.

Overall, the total cost for the MLE solution is given by the product of the cost for a single CG iteration and the total number of CG iteration. The number of CG iterations depends in turn on the noise level of the data, and the runtimes below are given for the full-sky and full-mission \Planck\ HFI 143\,GHz case. For a noise level of $\sigma = 200\, \mu$K, the algorithm requires 29 iterations to converge using the preconditioner from Eq.~\eqref{eq:preconditioner}, with a convergence criterion defined by a relative error of $\delta_{\rm{new}}/\delta_{0} = 10^{-10}$. Each iteration costs 32.9\,CPU-hrs, out of which 19.2\,CPU-hrs are used on parallel FFT calculations and application of the transfer function $T(\omega)$ within the $\tens{T}^{T} \tens{N}^{-1} \tens{T}$ operator. In order to parallelise and speed up the Fourier transformations efficiently, the TODs are divided into overlapping segments of length $2^{19}$, similar to \cite{planck2013-p03}. The run required a total of 563\,CPU-hrs, or 14.9 wall-hours when parallelized over 64 cores. Ultimately, this algorithm is intended to be integrated in the end-to-end CMB Gibbs sampler \texttt{Commander} \citep{BP03}. For comparison, the cost for a full Gibbs sample for all three \Planck\ LFI channels was 169\,CPU-hrs \citep{bp01,BP03,bp10}. Assuming no further optimizations, full analysis of all \Planck\ HFI channels with this algorithm will increase the total runtime by more than an order of magnitude. Full optimization will happen in the future Fortran implementation, and for now we simply conclude that this algorithm is indeed feasible, even though it is computationally expensive.


\section{Conclusions}
\label{sec:conclusions}

In this work we have proposed a new method for accounting for the finite bolometer transfer function in modern bolometer-based CMB experiments. This method integrates the bolometer transfer function directly into the classical maximum-likelihood mapmaking equation, which then is solved with a conjugate gradient method. This method is the optimal solution to the full deconvolution problem, and provides an unbiased signal estimate with proper noise weighting and correlations.

We have compared this method to the traditional two-step procedure used by most bolometer-based experiments to date, in which an estimate of the transfer function is deconvolved from the TOD prior to mapmaking. For slow detectors with a long bolometer time constant compared to the sampling rate, the deconvolution procedure boosts the white noise at high temporal frequencies, and this is usually regularized explicitly by an explicit and additional smoothing kernel. We have shown  that the resulting map estimate contains both significant scan-aligned residuals and larger noise correlations than the optimal method discussed in this paper. Calculating the effective ellipticity and FWHM of the beams resulting from the two deconvolution methods, we find that the ellipticity is $64\,\%$ lower for a \Planck\ 143\,GHz-based simulation with the MLE method than for the traditional method, and the FWHM is $2.3\,\%$ smaller. Another notable advantage of the optimal method is that it does not require an additional power spectrum level deconvolution kernel (because $\tens{K}=\tens{I}$ in this case), which should result in significantly lower beam estimation uncertainties when integrated into a full pipeline.



Based on these findings, we conclude that the new method is algebraically preferable over the traditional method. At the same time, the computational cost is also correspondingly higher, with a total runtime of many hundreds of CPU-hours for a typical \Planck\ HFI case. However, this cost will likely be decreased significantly both through better code optimization and algebraic improvements, for instance by implementing a better CG preconditioner. This is left for future work. In addition, the current implementation already results in runtimes that are fully feasible for modern computers. We anticipate that this method will allow for better data extraction both in reanalyses of the previous CMB experiments, such as \textit{Planck} HFI, and the analysis of the upcoming ones, such as Simons Observatory, LiteBIRD, and CMB-S4.

\begin{acknowledgements}
  The current work has received funding from the European Union’s
  Horizon 2020 research and innovation programme under grant agreement
  numbers 819478 (ERC; \textsc{Cosmoglobe}), 772253 (ERC;
  \textsc{bits2cosmology}), and 101007633 (Marie Skłodowska-Curie,
  \textsc{CMB-INFLATE}). In addition, the collaboration acknowledges
  support from RCN (Norway; grant no.\ 274990).  Some of the results
  in this paper have been derived using the \texttt{healpy} and
  \texttt{HEALPix}\footnote{\url{http://healpix.sf.net}} packages
  \citep{gorski2005, Zonca2019}.
\end{acknowledgements}

\bibliographystyle{aa}

\bibliography{refs,Planck_bib}

\begin{thebibliography}{27}
\expandafter\ifx\csname natexlab\endcsname\relax\def\natexlab#1{#1}\fi

\bibitem[{{Basyrov} {et~al.}(2023){Basyrov}, {Suur-Uski, A.-S.}, {Colombo, L.
  P. L.}, {Eskilt, J. R.}, {Paradiso, S.}, {Andersen, K. J.}, {Aurlien, R.},
  {Banerji, R.}, {Bersanelli, M.}, {Bertocco, S.}, {Brilenkov, M.}, {Carbone,
  M.}, {Eriksen, H. K.}, {Foss, M. K.}, {Franceschet, C.}, {Fuskeland, U.},
  {Galeotta, S.}, {Galloway, M.}, {Gerakakis, S.}, {Gjerl\o{}w, E.}, {Hensley,
  B.}, {Herman, D.}, {Iacobellis, M.}, {Ieronymaki, M.}, {Ihle, H. T.},
  {Jewell, J. B.}, {Karakci, A.}, {Keih\"anen, E.}, {Keskitalo, R.}, {Maggio,
  G.}, {Maino, D.}, {Maris, M.}, {Partridge, B.}, {Reinecke, M.}, {Svalheim, T.
  L.}, {Tavagnacco, D.}, {Thommesen, H.}, {Watts, D. J.}, {Wehus, I. K.}, \&
  {Zacchei, A.}}]{bp10}
{Basyrov}, A., {Suur-Uski, A.-S.}, {Colombo, L. P. L.}, {et~al.} 2023, A\&A,
  675, A10

\bibitem[{{Bennett} {et~al.}(1996){Bennett}, {Banday}, {Gorski}, {Hinshaw},
  {Jackson}, {Keegstra}, {Kogut}, {Smoot}, {Wilkinson}, \&
  {Wright}}]{COBE_4years}
{Bennett}, C.~L., {Banday}, A.~J., {Gorski}, K.~M., {et~al.} 1996, \apjl, 464,
  L1

\bibitem[{{Bersanelli} {et~al.}(2010){Bersanelli}, {Mandolesi}, {Butler},
  {Mennella}, {Villa}, {Aja}, {Artal}, {Artina}, {Baccigalupi}, {Balasini},
  {Baldan}, {Banday}, {Bastia}, {Battaglia}, {Bernardino}, {Blackhurst},
  {Boschini}, {Burigana}, {Cafagna}, {Cappellini}, {Cavaliere}, {Colombo},
  {Crone}, {Cuttaia}, {D'Arcangelo}, {Danese}, {Davies}, {Davis}, {de Angelis},
  {de Gasperis}, {de La Fuente}, {de Rosa}, {de Zotti}, {Falvella}, {Ferrari},
  {Ferretti}, {Figini}, {Fogliani}, {Franceschet}, {Franceschi}, {Gaier},
  {Garavaglia}, {Gomez}, {Gorski}, {Gregorio}, {Guzzi}, {Herreros},
  {Hildebrandt}, {Hoyland}, {Hughes}, {Janssen}, {Jukkala}, {Kettle},
  {Kilpi{\"a}}, {Laaninen}, {Lapolla}, {Lawrence}, {Lawson}, {Leahy},
  {Leonardi}, {Leutenegger}, {Levin}, {Lilje}, {Lowe}, {Lubin}, {Maino},
  {Malaspina}, {Maris}, {Marti-Canales}, {Martinez-Gonzalez}, {Mediavilla},
  {Meinhold}, {Miccolis}, {Morgante}, {Natoli}, {Nesti}, {Pagan}, {Paine},
  {Partridge}, {Pascual}, {Pasian}, {Pearson}, {Pecora}, {Perrotta},
  {Platania}, {Pospieszalski}, {Poutanen}, {Prina}, {Rebolo}, {Roddis},
  {Rubi{\~n}o-Martin}, {Salmon}, {Sandri}, {Seiffert}, {Silvestri},
  {Simonetto}, {Sjoman}, {Smoot}, {Sozzi}, {Stringhetti}, {Taddei}, {Tauber},
  {Terenzi}, {Tomasi}, {Tuovinen}, {Valenziano}, {Varis}, {Vittorio}, {Wade},
  {Wilkinson}, {Winder}, {Zacchei}, \& {Zonca}}]{bersanelli2010}
{Bersanelli}, M., {Mandolesi}, N., {Butler}, R.~C., {et~al.} 2010, \aap, 520,
  A4

\bibitem[{{BeyondPlanck Collaboration}(2023)}]{bp01}
{BeyondPlanck Collaboration}. 2023, A\&A, 675, A1

\bibitem[{{Fosalba} {et~al.}(2002){Fosalba}, {Dor{\'e}}, \&
  {Bouchet}}]{Fosalba_2002}
{Fosalba}, P., {Dor{\'e}}, O., \& {Bouchet}, F.~R. 2002, \prd, 65, 063003

\bibitem[{{Galloway} {et~al.}(2023){Galloway}, {Andersen}, {Aurlien},
  {Banerji}, {Bersanelli}, {Bertocco}, {Brilenkov}, {Carbone}, {Colombo},
  {Eriksen}, {Eskilt}, {Foss}, {Franceschet}, {Fuskeland}, {Galeotta},
  {Gerakakis}, {Gjerl{\o}w}, {Hensley}, {Herman}, {Iacobellis}, {Ieronymaki},
  {Ihle}, {Jewell}, {Karakci}, {Keih{\"a}nen}, {Keskitalo}, {Maggio}, {Maino},
  {Maris}, {Mennella}, {Paradiso}, {Partridge}, {Reinecke}, {San}, {Suur-Uski},
  {Svalheim}, {Tavagnacco}, {Thommesen}, {Watts}, {Wehus}, \& {Zacchei}}]{BP03}
{Galloway}, M., {Andersen}, K.~J., {Aurlien}, R., {et~al.} 2023, \aap, 675, A3

\bibitem[{{G{\'o}rski} {et~al.}(2005){G{\'o}rski}, {Hivon}, {Banday},
  {Wandelt}, {Hansen}, {Reinecke}, \& {Bartelmann}}]{gorski2005}
{G{\'o}rski}, K.~M., {Hivon}, E., {Banday}, A.~J., {et~al.} 2005, \apj, 622,
  759

\bibitem[{Harris {et~al.}(2020)Harris, Millman, van~der Walt, Gommers,
  Virtanen, Cournapeau, Wieser, Taylor, Berg, Smith, Kern, Picus, Hoyer, van
  Kerkwijk, Brett, Haldane, del R{\'{i}}o, Wiebe, Peterson,
  G{\'{e}}rard-Marchant, Sheppard, Reddy, Weckesser, Abbasi, Gohlke, \&
  Oliphant}]{numpy}
Harris, C.~R., Millman, K.~J., van~der Walt, S.~J., {et~al.} 2020, Nature, 585,
  357

\bibitem[{{Hinshaw} {et~al.}(2013){Hinshaw}, {Larson}, {Komatsu}, {Spergel},
  {Bennett}, {Dunkley}, {Nolta}, {Halpern}, {Hill}, {Odegard}, {Page}, {Smith},
  {Weiland}, {Gold}, {Jarosik}, {Kogut}, {Limon}, {Meyer}, {Tucker}, {Wollack},
  \& {Wright}}]{WMAP_9years_cosm}
{Hinshaw}, G., {Larson}, D., {Komatsu}, E., {et~al.} 2013, \apjs, 208, 19

\bibitem[{Howlett {et~al.}(2012)Howlett, Lewis, Hall, \&
  Challinor}]{Howlett_2012}
Howlett, C., Lewis, A., Hall, A., \& Challinor, A. 2012, \jcap, 1204, 027

\bibitem[{{Ihle} {et~al.}(2023){Ihle}, {Bersanelli}, {Franceschet},
  {Gjerl{\o}w}, {Andersen}, {Aurlien}, {Banerji}, {Bertocco}, {Brilenkov},
  {Carbone}, {Colombo}, {Eriksen}, {Eskilt}, {Foss}, {Fuskeland}, {Galeotta},
  {Galloway}, {Gerakakis}, {Hensley}, {Herman}, {Iacobellis}, {Ieronymaki},
  {Jewell}, {Karakci}, {Keih{\"a}nen}, {Keskitalo}, {Maggio}, {Maino}, {Maris},
  {Mennella}, {Paradiso}, {Partridge}, {Reinecke}, {San}, {Suur-Uski},
  {Svalheim}, {Tavagnacco}, {Thommesen}, {Watts}, {Wehus}, \&
  {Zacchei}}]{BP_VI}
{Ihle}, H.~T., {Bersanelli}, M., {Franceschet}, C., {et~al.} 2023, \aap, 675,
  A6

\bibitem[{{Keih{\"a}nen} \& {Reinecke}(2012)}]{ArtDeco}
{Keih{\"a}nen}, E. \& {Reinecke}, M. 2012, \aap, 548, A110

\bibitem[{{Lamarre} {et~al.}(2010){Lamarre}, {Puget}, {Ade}, {Bouchet},
  {Guyot}, {Lange}, {Pajot}, {Arondel}, {Benabed}, {Beney}, {Beno{\^i}t},
  {Bernard}, {Bhatia}, {Blanc}, {Bock}, {Br{\'e}elle}, {Bradshaw}, {Camus},
  {Catalano}, {Charra}, {Charra}, {Church}, {Couchot}, {Coulais}, {Crill},
  {Crook}, {Dassas}, {de Bernardis}, {Delabrouille}, {de Marcillac}, {Delouis},
  {D{\'e}sert}, {Dumesnil}, {Dupac}, {Efstathiou}, {Eng}, {Evesque},
  {Fourmond}, {Ganga}, {Giard}, {Gispert}, {Guglielmi}, {Haissinski},
  {Henrot-Versill{\'e}}, {Hivon}, {Holmes}, {Jones}, {Koch}, {Lagard{\`e}re},
  {Lami}, {Land{\'e}}, {Leriche}, {Leroy}, {Longval},
  {Mac{\'{\i}}as-P{\'e}rez}, {Maciaszek}, {Maffei}, {Mansoux}, {Marty}, {Masi},
  {Mercier}, {Miville-Desch{\^e}nes}, {Moneti}, {Montier}, {Murphy},
  {Narbonne}, {Nexon}, {Paine}, {Pahn}, {Perdereau}, {Piacentini}, {Piat},
  {Plaszczynski}, {Pointecouteau}, {Pons}, {Ponthieu}, {Prunet}, {Rambaud},
  {Recouvreur}, {Renault}, {Ristorcelli}, {Rosset}, {Santos}, {Savini},
  {Serra}, {Stassi}, {Sudiwala}, {Sygnet}, {Tauber}, {Torre}, {Tristram},
  {Vibert}, {Woodcraft}, {Yurchenko}, \& {Yvon}}]{lamarre2010}
{Lamarre}, J., {Puget}, J., {Ade}, P.~A.~R., {et~al.} 2010, \aap, 520, A9

\bibitem[{Lewis {et~al.}(2000)Lewis, Challinor, \& Lasenby}]{Lewis_1999}
Lewis, A., Challinor, A., \& Lasenby, A. 2000, \apj, 538, 473

\bibitem[{{\sorthelp{Planck Collaboration 2014A}}{Planck Collaboration
  I}(2014)}]{planck2013-p01}
{\sorthelp{Planck Collaboration 2014A}}{Planck Collaboration I}. 2014, \aap,
  571, A1

\bibitem[{{\sorthelp{Planck Collaboration 2014F}}{Planck Collaboration
  VI}(2014)}]{planck2013-p03}
{\sorthelp{Planck Collaboration 2014F}}{Planck Collaboration VI}. 2014, \aap,
  571, A6

\bibitem[{{\sorthelp{Planck Collaboration 2014G}}{Planck Collaboration
  VII}(2014)}]{planck2013-p03c}
{\sorthelp{Planck Collaboration 2014G}}{Planck Collaboration VII}. 2014, \aap,
  571, A7

\bibitem[{{\sorthelp{Planck Collaboration 2015A}}{Planck Collaboration
  I}(2016)}]{planck2014-a01}
{\sorthelp{Planck Collaboration 2015A}}{Planck Collaboration I}. 2016, \aap,
  594, A1

\bibitem[{{\sorthelp{Planck Collaboration 2015G}}{Planck Collaboration
  VII}(2016)}]{planck2014-a08}
{\sorthelp{Planck Collaboration 2015G}}{Planck Collaboration VII}. 2016, \aap,
  594, A7

\bibitem[{{\sorthelp{Planck Collaboration 2018A}}{Planck Collaboration
  I}(2020)}]{planck2016-l01}
{\sorthelp{Planck Collaboration 2018A}}{Planck Collaboration I}. 2020, \aap,
  641, A1

\bibitem[{{\sorthelp{Planck Collaboration 2018E}}{Planck Collaboration
  V}(2020)}]{planck2016-l05}
{\sorthelp{Planck Collaboration 2018E}}{Planck Collaboration V}. 2020, \aap,
  641, A5

\bibitem[{Shewchuk(1994)}]{CG_pain}
Shewchuk, J.~R. 1994, An Introduction to the Conjugate Gradient Method Without
  the Agonizing Pain, Edition $1\frac14$,
  \href{http://www.cs.cmu.edu/~quake-papers/painless-conjugate-gradient.pdf}{http://www.cs.cmu.edu/\textasciitilde{}quake-papers/painless-conjugate-gradient.pdf}

\bibitem[{{Stevens} {et~al.}(2020){Stevens}, {Cothard}, {Vavagiakis}, {Ali},
  {Arnold}, {Austermann}, {Choi}, {Dober}, {Duell}, {Duff}, {Hilton}, {Ho},
  {Hoang}, {Hubmayr}, {Lee}, {Mangu}, {Nati}, {Niemack}, {Raum}, {Renzullo},
  {Salatino}, {Sasse}, {Simon}, {Staggs}, {Suzuki}, {Truitt}, {Ullom},
  {Vivalda}, {Vissers}, {Walker}, {Westbrook}, {Wollack}, {Xu}, \&
  {Yohannes}}]{Stevens_2020}
{Stevens}, J.~R., {Cothard}, N.~F., {Vavagiakis}, E.~M., {et~al.} 2020, Journal
  of Low Temperature Physics, 199, 672

\bibitem[{{Tegmark}(1997)}]{tegmark_1997}
{Tegmark}, M. 1997, \apjl, 480, L87

\bibitem[{Virtanen {et~al.}(2020)Virtanen, Gommers, Oliphant, Haberland, Reddy,
  Cournapeau, Burovski, Peterson, Weckesser, Bright, {van der Walt}, Brett,
  Wilson, Millman, Mayorov, Nelson, Jones, Kern, Larson, Carey, Polat, Feng,
  Moore, {VanderPlas}, Laxalde, Perktold, Cimrman, Henriksen, Quintero, Harris,
  Archibald, Ribeiro, Pedregosa, {van Mulbregt}, \& {SciPy 1.0
  Contributors}}]{Scipy}
Virtanen, P., Gommers, R., Oliphant, T.~E., {et~al.} 2020, Nature Methods, 17,
  261

\bibitem[{{Zhao} {et~al.}(2008){Zhao}, {Allen}, {Amiri}, {Appel},
  {Battistelli}, {Burger}, {Chervenak}, {Dahlen}, {Denny}, {Devlin}, {Dicker},
  {Doriese}, {D{\"u}nner}, {Essinger-Hileman}, {Fisher}, {Fowler}, {Halpern},
  {Hilton}, {Hincks}, {Irwin}, {Jarosik}, {Klein}, {Lau}, {Marriage},
  {Martocci}, {Moseley}, {Niemack}, {Page}, {Parker}, {Sederberg}, {Staggs},
  {Stryzak}, {Swetz}, {Switzer}, {Thornton}, \& {Wollack}}]{Zhao_2008}
{Zhao}, Y., {Allen}, C., {Amiri}, M., {et~al.} 2008, in Society of
  Photo-Optical Instrumentation Engineers (SPIE) Conference Series, Vol. 7020,
  Millimeter and Submillimeter Detectors and Instrumentation for Astronomy IV,
  ed. W.~D. {Duncan}, W.~S. {Holland}, S.~{Withington}, \& J.~{Zmuidzinas},
  70200O

\bibitem[{Zonca {et~al.}(2019)Zonca, Singer, Lenz, Reinecke, Rosset, Hivon, \&
  Gorski}]{Zonca2019}
Zonca, A., Singer, L.~P., Lenz, D., {et~al.} 2019, Journal of Open Source
  Software, 4, 1298

\end{thebibliography}

\end{document}